\documentclass[12pt]{article}
\usepackage{graphicx}

\renewcommand{\abstract}[1]{{ \footnotesize \noindent {\bf Abstract} #1 \\}}
\renewcommand{\author}[1]{\subsubsection*{\bf#1}}
\newcommand{\address}[1]{\subsubsection*{\it#1}}

\newcommand{\chapter}[1]{{\Large \bf \noindent #1}}

\begin{document}

\chapter{Mass modelling of Abell 2634: avoiding the interloper bias}

\author{R. Wojtak and E. L. {\L}okas}
\address{Nicolaus Copernicus Astronomical Center, Bartycka 18, 00-716 Warsaw, Poland}
\abstract{Using an example of the Abell 2634 galaxy cluster we discuss the effect of
contamination of kinematic data by interlopers and its impact on mass modelling. The cluster
data reveal
rich substructure along the line of sight. We demonstrate that it is necessary to apply
a few independent methods of interloper removal in order to obtain a reliable sample
of cluster members. We present results of three such methods which are
commonly used in the literature and  have been recently extensively tested on
simulated data. Only two of them lead to consistent and reliable samples of cluster galaxies.
For both of them we provide parameters of the best-fitting NFW density profile
by fitting an isotropic solution of the Jeans equation to the velocity dispersion profiles.}

\section{The Abell 2634 galaxy cluster}

\begin{figure}\label{3methods}
\begin{center}
\leavevmode
\includegraphics[width=1.\textwidth]{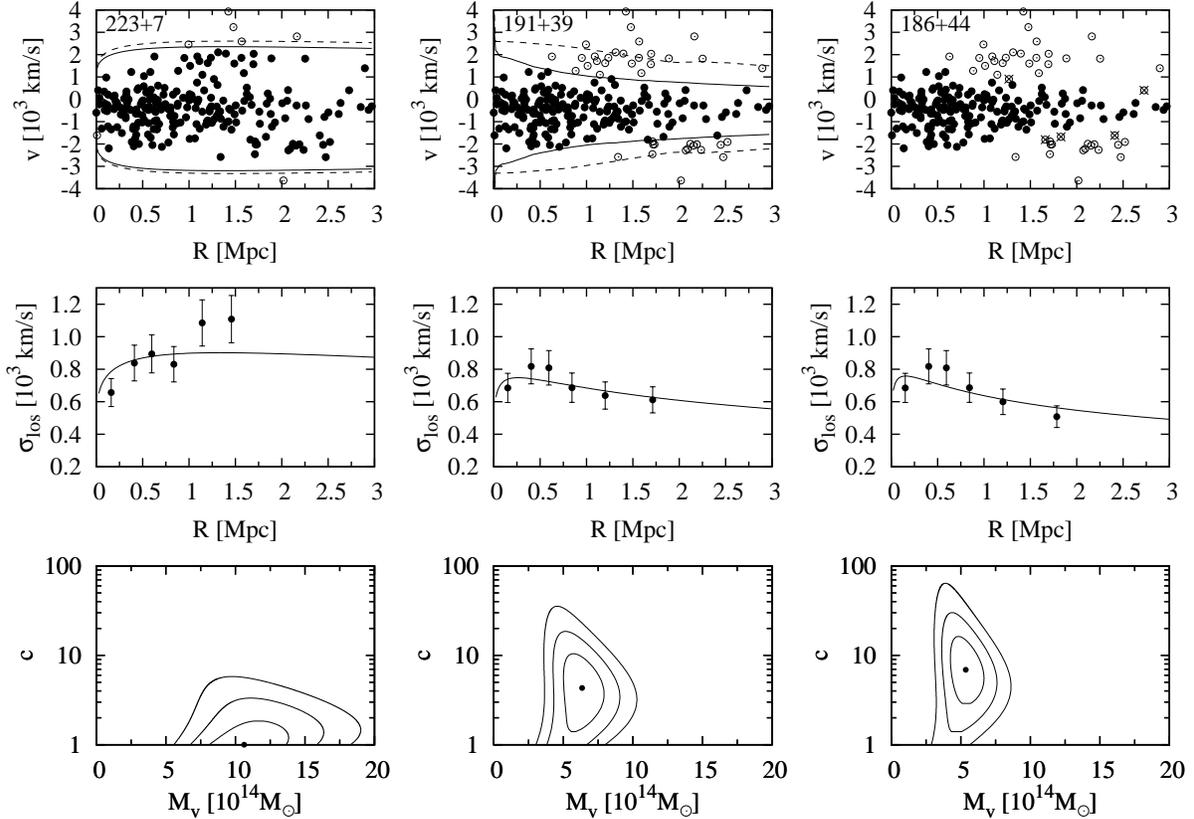}
\caption{Results of interloper removal applied to the kinematic data for A2634.
Columns from the left to the right correspond to the $3\sigma_{\rm los}(R)$, $v_{\rm max}(R)$
and $M_{P}/M_{VT}$ methods. Open and filled symbols in the velocity diagrams (upper row)
stand for the identified interlopers and cluster members respectively 
(the numbers of points of both groups are given in the upper left corners). 
Dashed and solid lines
plotted there indicate velocity envelopes from the first and the last
iteration of a given method. The middle row shows the dispersion profiles corresponding
to the different samples of cluster galaxies and the best-fitting solution of the
Jeans equation for isotropic orbits (solid lines). Results of the fitting in terms of
$1\sigma$, $2\sigma$ and $3\sigma$
confidence regions in the $M_v-c$ parameter plane are shown in the bottom row.}
\end{center}
\end{figure}

Abell 2634 is a nearby galaxy cluster at redshift $z=0.0314$. At present there are
$230$ galaxies with measured spectroscopic redshift available within the aperture
of $3$ Mpc around the cluster centre. The data, extracted from the NASA/IPAC Extragalactic
Database (NED), are presented in the upper row of Figure~\ref{3methods}
in the form of the so-called velocity diagram, in which velocities $v$ of galaxies
in the rest frame of the cluster are plotted against projected distances $R$ from
the cluster centre.

The cluster has a relatively regular image in X-rays and possesses a Bright
Central Galaxy whose position coincides with the global maximum of the
surface distribution of the X-ray emitting gas. This means
that the object is expected to be nearly in dynamical equilibrium and all
methods of mass inference involving assumptions of relaxation can be reliably
applied. However, the
distribution of galaxies in the velocity diagram is not as
regular as one could expect from this picture. It reveals two groups of
interlopers (background and foreground galaxies seen due to projection
effects) probably due to the presence of two structures gravitationally
unbound to the cluster but positioned along the line of sight. This kind of data
contamination may lead to significant systematic errors in mass estimates.
Reliable inferences concerning the mass distribution thus require careful
selection of cluster members.

\section{Removal of interlopers}

No method exists which would allow us to discriminate between the cluster
members and interlopers with certainty. Any galaxy observed in the direction
of a cluster can belong to this system as well as be seen in the sample only
due to projection effects. However, the probability of observing a cluster galaxy
is not the same for all objects: galaxies with small velocities in the rest
frame of the cluster belong physically to the system with more probability than
high-velocity galaxies. Therefore the key idea to handle the problem of interloper
identification is to introduce an envelope $\pm v_{\rm lim}(R)$ around the systemic
velocity of the cluster,
which separates the area of the velocity diagram preferentially occupied by cluster
members. Then the galaxies with velocities exceeding the
boundary line $\pm v_{\rm lim}(R)$ are
suspected to be interlopers and should be rejected from the sample.

As shown by Wojtak et al. (2007), for the best methods the rate of identification
of gravitationally unbound galaxies reaches the efficiency of 60-70$\%$ (for
the initial cut-off in velocity of about 3-4$\sigma_{\rm los}$), which is
close to the maximum value available. On the other hand, the final data sample
still possesses some residual contamination of background galaxies.
Moreover, some cluster members are lost due to unavoidable uncertainty of each method.
However, fractions of both types of galaxies, i.e. lost cluster members and remaining
interlopers, are of the order of 2-3$\%$ and have negligible impact on the final
estimate of the mass profile. In the following we present three typical
methods of interloper removal, which were found to work well on simulated
as well as real data (for details see {\L}okas et al. 2006,
Wojtak et al. 2007 and Wojtak \& {\L}okas 2007),
and apply them to the data for A2634.

\begin{figure}\label{mvt}
\centering
\includegraphics[width=0.5\textwidth]{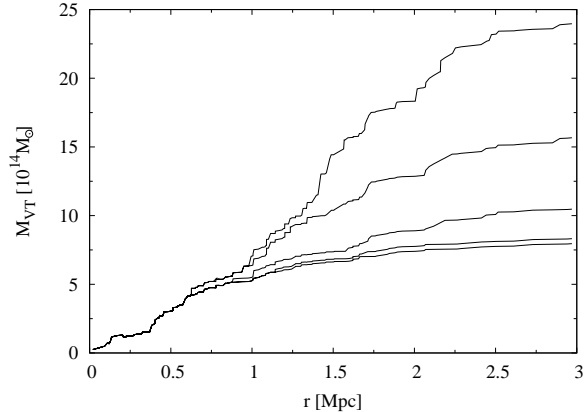}
\caption{Profiles of the virial mass estimator obtained in the subsequent iterations of the
$v_{\rm max}(R)$ method.}
\end{figure}

In the first, rather natural approach (hereafter $3\sigma_{\rm los}(R)$ method) velocity
envelope is determined by
\begin{equation}\label{3sigma}
	v_{\rm lim}(R) = 3\sigma_{\rm los}(R),
\end{equation}
where $\sigma_{\rm los}(R)$ is the line-of-sight velocity dispersion profile
given by the projected solution of the Jeans equation for isotropic orbits
(the anisotropy parameter $\beta=0$)
fitted to the data (see {\L}okas et al. 2001 and {\L}okas et al. 2006 for details).
This formula is a generalization of the original concept by Yahil \& Vidal (1977),
who proposed to eliminate interlopers via $3\sigma_{\rm los}$ clipping where
$\sigma_{\rm los}$ was the total velocity dispersion of the cluster.
The interlopers are removed
iteratively in a sequence of dispersion fitting and rejection of outliers.
The results of this method applied
to the data of A2634 are shown in the upper left panel of Figure~\ref{3methods}.
The method identified and rejected only $7$
interlopers marked with open symbols
(the 7th rejected galaxy in the very centre is probably a member).

The next approach (labelled hereafter $v_{\rm max}(R)$) was originally
proposed by den Hartog \& Katgert (1996). Considering
boundary conditions for the velocity field in virialized objects, they showed that the velocity
envelope could be well approximated by the maximum of the projection along the line of sight
of the circular velocity vector ${\mathbf v}_{\rm cir}$ and the infall velocity vector
${\mathbf v}_{\rm inf}=\sqrt{2}{\mathbf v}_{\rm cir}$, namely:
\begin{equation}\label{vmax}
	v_{\rm lim}(R)=
	{\rm max}_{R}\{v_{\rm cir}\sin\theta,
	v_{\rm inf}\cos\theta\},
\end{equation}
where $\theta$ is the angle between the position vector with respect to the cluster centre and
the line of sight. The mass which is required to calculate the circular velocity
$v_{\rm cir}=[GM(r)/r]^{1/2}$, is approximated by the mass estimator based on virial theorem
$M_{VT}(R)$
\begin{equation}\label{M_VT}
	M(r)\approx M_{VT}(R=r)=\frac{3\pi N}{2G}
	\frac{\Sigma_{i}(v_{i}-\langle v\rangle)^{2}}{\Sigma_{i>j}1/R_{i,j}},
\end{equation}
where $R_{i,j}$ is the projected distance between the $i$-th and the $j$-th galaxy and $N$
is the number of galaxies in the sample. Inserting this formula into equation (\ref{vmax}) one
can evaluate the $v_{\rm lim}(R)$ profile numerically. Again, interlopers are removed from the data
iteratively until all galaxies are enclosed within the velocity envelope.

The velocity diagram in the upper middle panel of Figure~\ref{3methods}
shows the $\pm v_{\rm lim}(R)$ profiles of the first
(dashed line) and final (solid line) iteration of this method applied to the
data for A2634. The final sample of cluster members ($191$ galaxies) is indicated
with filled circles, whereas interlopers ($39$ galaxies) with open symbols.
Mass profiles obtained in subsequent iterations of the method are plotted in Figure~\ref{mvt}
with the highest line for the first iteration and the lowest one for the last.
One can see how succeeding steps of the procedure eliminate mass overestimation
caused by the contamination of the galaxy sample by gravitationally unbound objects.

\begin{figure}\label{jack}
\centering
\includegraphics[width=1.0\textwidth]{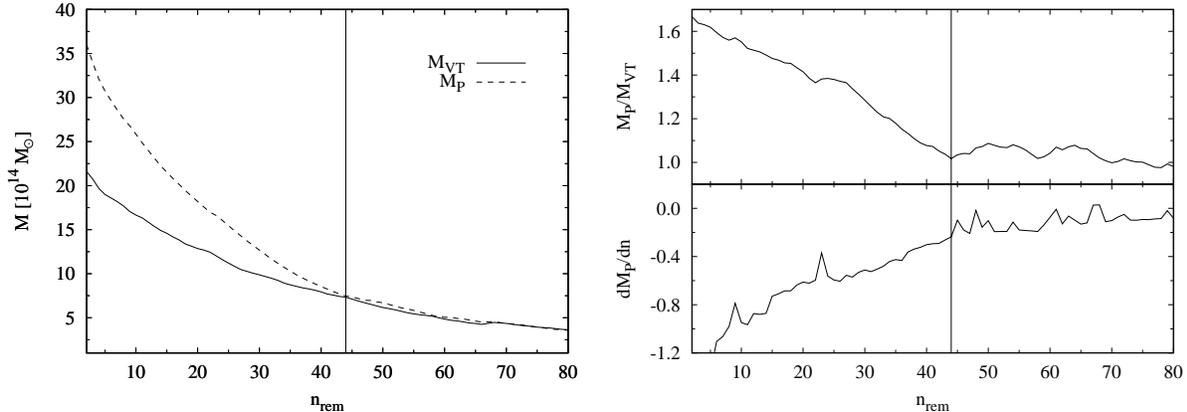}
\caption{The mass estimators $M_{VT}$ and $M_{P}$, the ratio $M_{P}/M_{VT}$ and
${\rm d} M_{P}/{\rm d}n$ as functions of the number $n_{\rm rem}$ of galaxies removed from the sample by the jackknife technique. In each panel the vertical solid line indicates the 
moment of convergence $n_{\rm rem}=44$, when the number of eliminated interlopers is probably largest.}
\end{figure}

The third method which proves rather efficient in removing the interlopers
(labelled hereafter $M_{P}/M_{VT}$), is based on the analysis of the relative bias
of two mass estimators, the virial mass estimator $M_{VT}$ defined by (\ref{M_VT})
and the so-called projected mass estimator $M_{P}$ given by
\begin{equation}\label{M_{P}}
	M_{P}=\frac{32}{\pi GN}\sum_{i}(v_{i}-\langle v\rangle)^{2}R_{i}.
\end{equation}
As shown by Perea et al. (1990), the interlopers are the main source of
bias of both mass estimators and one can use the jackknife statistics to
eliminate them from the cluster sample. The moment of convergence could
be found by investigating $M_{P}/M_{VT}$ and ${\rm d} M/{\rm d}n$ profiles as
functions of the number of removed galaxies $n_{\rm rem}$. As demonstrated
by Wojtak et al. (2007), the maximum number of removed interlopers is
signified by a characteristic knee-like point between the rapidly
varying part of the $M_{P}/M_{VT}$ profile and its plateau extension.
Figure~\ref{jack} shows the detection of this point in the case of A2634. The
$n_{\rm rem}=44$ interlopers removed in this procedure are indicated
with open circles in the velocity diagram shown in the upper right panel
of Figure~\ref{3methods}. The method identifies $5$ more interlopers
(marked with crosses) than the $v_{\rm max}(R)$ approach.

\section{Conclusions}

A2634 is an example of a cluster with kinematic sample significantly
contaminated by interlopers. As discussed by Wojtak \& {\L}okas
(2007), for such objects the $3\sigma_{\rm los}(R)$ method converges too early
and leaves in the sample a non-negligible number of gravitationally unbound galaxies. On
the other hand, the $v_{\rm max}(R)$ and $M_{P}/M_{VT}$ methods lead to much more
consistent and reliable samples of cluster members. This fact manifests itself also in terms
of mass modelling. The middle row of Figure~\ref{3methods} shows the
dispersion profiles of the corresponding samples of cluster members. The
bottom panels demonstrate results of fitting an isotropic solution
of the Jeans equation with NFW density profile (Navarro, Frenk \& White 1997)
involving two free parameters, the virial mass $M_{v}$ and concentration $c$.

As expected, the result for
the sample obtained with the $3\sigma_{\rm los}(R)$ method still reveals
features characteristic for interloper bias, i.e. the inferred density profile
is flat $(c=1)$ and the mass is significantly overestimated with respect to the
results obtained for samples from other interloper removal schemes.
On the other hand, confidence contours obtained for the other two samples
agree well with each other. This indicates that these two samples of
cluster members are reliable. The best-fitting parameters for the $v_{\rm max}(R)$
method are: $M_{v}=6.4^{+1.5}_{-1.4}\times 10^{14}M_{\odot}$
and $c=4.3^{+6.1}_{-2.9}$ (1$\sigma$ errors) while for the $M_{P}/M_{VT}$ method we find
$M_{v}=5.4^{+1.3}_{-1.2}\times 10^{14}M_{\odot}$ and $c=6.9^{+9.3}_{-4.0}$.
Note that when the assumption of isotropy is relaxed and kurtosis is added to the
analysis to constrain all three parameters, $M_v$, $c$ and anisotropy $\beta$, the
result is not very different, i.e. still isotropic orbits are preferred by the data
(see Wojtak \& {\L}okas 2007).



\begin{thebibliography}{}

\bibitem[1]{hartog} den Hartog R., Katgert P., 1996, MNRAS, 279, 349
\bibitem[2]{nav} Navarro J. F., Frenk C. S., White S. D. M., 1997, ApJ, 490, 493
\bibitem[3]{lok1} {\L}okas E. L., Mamon G. A., 2001, MNRAS, 321, 155
\bibitem[4]{perea} Perea J., del Olmo A., Moles M., 1990, A\&A, 237, 319
\bibitem[5]{yah} Yahil A., Vidal N. V., 1977, ApJ, 214, 347
\bibitem[6]{woj1} Wojtak R., {\L}okas E. L., Mamon G. A., Gottl\"ober S.,
Prada F., Moles M., 2007, A\&A, 466, 437
\bibitem[7]{woj2} Wojtak R., {\L}okas E. L., 2007, MNRAS, 377, 843
\end{thebibliography}
\end{document}